\documentclass[prl,twocolumn,showpacs,preprintnumbers,superscriptaddress]{revtex4}

\usepackage{times}
\usepackage{subfig}
\usepackage{bm}
\usepackage{graphicx}
\usepackage{amsbsy}
\usepackage{amsmath}
\usepackage{amsfonts}
\usepackage{amsthm}
\usepackage{float}
\usepackage{color}

\begin{document}
 \theoremstyle{plain}
\newtheorem{theorem}{Theorem}
\newtheorem{lemma}[theorem]{Lemma}
\newtheorem{corollary}[theorem]{Corollary}
\newtheorem{proposition}[theorem]{Proposition}
\newtheorem{conjecture}[theorem]{Conjecture}

\theoremstyle{definition}
\newtheorem{definition}[theorem]{Definition}

\title{Locally Unextendible Non-Maximally
Entangled Basis}
\author {Indranil Chakrabarty}
\affiliation{Institute of Physics, Sainik School Post, 
Bhubaneswar-751005, Orissa, India}
\author{Pankaj Agrawal}
\affiliation{Institute of Physics, Sainik School Post, 
Bhubaneswar-751005, Orissa, India}
\author {Arun K Pati}
\affiliation{Harish Chandra Research Institute, Chhatnag Road, Jhunsi, Allahabad-211019, UP, India}

\begin{abstract}
We introduce the concept of the locally unextendible non-maximally entangled basis (LUNMEB) in $H^d \bigotimes H^d$.
It is shown that such a basis consists
of $d$ orthogonal vectors for a non-maximally entangled state. However, there can be a maximum of  $(d-1)^2$ orthogonal
vectors for non-maximally entangled state if it is maximally entangled in $(d-1)$ dimensional subspace. Such a basis plays an 
important role in determining the number of classical bits that one can send in a superdense coding protocol
 using a non-maximally entangled state as a resource. By constructing appropriate POVM  operators,
 we find that the number of classical bits one can transmit using a non-maximally entangled
state as a resource is $(1+p_0\frac{d}{d-1})\log d$, where $p_0$ is the smallest Schmidt coefficient. 
However, when the state is maximally entangled in its subspace then one can send up to $2\log (d-1) $ bits.
We also find that for $d= 3$, former may be more suitable for the superdense coding.
\end{abstract}

\maketitle
\section{Introduction}
It is the Einstein, Podolsky, and Rosen (EPR) paper where for the first time
 entangled states \cite{alb} were used to explore the mysterious nature of the formalism
 of quantum mechanics.  In the course of time, it has been recognized by the scientific community that the difference
between factorisable and entangled (non-factorisable) quantum states is pivotal in understanding the deepest nature of reality. In 1964, Bell was the first person to show that this entanglement implies lack of local realism in
quantum mechanics \cite{bell}. It was quite surprising when
it was found that there are sets of product states which
nevertheless display a form of nonlocality \cite{divi, ben1}. It
was shown that there are sets of orthogonal product vectors $S$ (say) of a tensor product Hilbert space $H^n
\bigotimes H^m$ $(n,m > 2)$
such that even in the complementary set there are product states which are orthogonal
to every state in the set $S$. However, we will never be able to find enough states so as to complete the set to form a full basis of the Hilbert space $H^n \bigotimes H^m$ $(n,m > 2)$. Such a basis is called an Uncompletable Product Basis (UCPB) \cite{divi}. It is called Unextendible Product Basis (UPB) if it is not possible to find at least one 
 product vector in the complement of the set S which is orthogonal to all the members of the set $S$ \cite{ben1, horo1, horo2, chatur, pitt, brav, cohen}. The notion of unextendibility gives
rise to two important quantum phenomena. (i) The mixed state which lies on
the subspace complementary to the subspace spanned by UPB is a bound entangled
state \cite{horo1,horo2}. It is bound in the sense that no free 
entanglement can be distilled from it. In reference \cite{ben1}, a systematic way of constructing bound entangled states had been provided for the first time. (ii) The states
comprising a UPB are locally immeasurable \cite{ben2}, i.e. an
unknown member of the set cannot be reliably distinguished
from the others by applying local measurements and by communicating classically.
Recently, the notion of unextendible maximally entangled basis (UMEB) in a
restricted situation $ H^d \bigotimes H^d$ has been introduced. The basis set consists of fewer than $d^2$ vectors and has
no additional maximally entangled vectors orthogonal to all of them. It was shown that UMEBs do not exist for d = 2 and there exists a 6-member and 12- member UMEB for three and four dimensional cases \cite{bra}.
In this paper,  we introduce the notion of locally unextendible non-maximally entangled basis (LUNMEB) in $ H^d \bigotimes H^d$. We show that if we start with a non-maximally entangled state in $H^d \bigotimes H^d$ and if one party applies local unitary operations, we will have $d^2$ vectors out of which we can build up $d$ classes. Each class has $d$  non-maximally entangled vectors, in such a way that they are mutually orthogonal to each other. We show that there does not exist any local unitary transformation that will create a non-maximally
entangled vector  orthogonal to each of these $d$ orthogonal states. So, these $d$ orthogonal vectors will form a basis which is unextendible in the sense that they can not be extended locally. 
Let us first of all give the formal definition of locally unextendible non-maximally entangled basis (LUNMEB).

\noindent
\textbf{Definition:} A set of states $\{|\psi_a\rangle\in H^d \bigotimes H^d\}, a=1,2,..n$ is called LUNMEB iff\\
(i) all states $|\psi_a\rangle$ are non-maximally entangled.\\
(ii)$\langle \psi_a|\psi_b\rangle=\delta_{a,b}$.\\
(iii) For all $a$ and $b$, there exists local unitary transformation $U_{ba}$ such that $(U_{ba}\bigotimes I)|\psi_a\rangle=|\psi_b\rangle$.\\ 
(iv) If $\langle \psi_a|\psi\rangle=0, \forall a=1,2,..,n$, then there exists no unitary transformation $U$, such that  $(U\bigotimes I)|\psi_a\rangle=|\psi\rangle,\forall a=1,2,..,n$.\\

One can in principle construct an unextendible non-maximally entangled basis differently and it can have more than $d$ orthogonal vectors. The importance of considering the restricted class of locally unextendible non-maximally entangled basis lies in the context of super dense coding. In super dense coding, we create the orthogonal vectors by applying local unitaries. In this context, it is interesting to study those vectors which are  local unitary equivalent.
One can then answer the question: given an non-maximally entangled resource, what is the amount of classical information one can communicate.  We will see that this locally unextendible non-maximally entangled basis (LUNMEB) is going to play an important role in determining the number of bits that one can communicate through a super dense coding protocol. 

The organization of the paper is as follows. In section 2 and 3, we construct locally unextendible non-maximally entangled basis (LUNMEB) for
$H^2 \bigotimes H^2$ and $H^3 \bigotimes H^3$ system. In section 4,  we give the most general construction for $H^d \bigotimes H^d$ systems. In section 5, we construct a set of POVM operators for a genuinely non-maximally entangled set of vectors.
In the next section, we discuss the importance of  LUNMEB in the context of super dense coding. Finally, we conclude in the last section. 

\section{Locally Unextendible Non-Maximally Entangled Basis in $H^2 \bigotimes H^2$ }

Let us first of all consider a non-maximally entangled state  $|\phi_{2}\rangle$ in $ H^2 \bigotimes H^2$, explicitly written in the 
Schmidt decomposition form as
\begin{eqnarray}
  |\phi_{2}\rangle=\sum_{k=0}^{1}C_k|kk\rangle=C_0|0\rangle|0\rangle+C_1|1\rangle|1\rangle.
 \end{eqnarray}
We show that this state gives rise to an unextendible non-maximally entangled basis of dimension two under local unitary operations. Let us consider a set of local unitary operators, in $d$ dimension, which itself forms a basis
\begin{eqnarray}
 U_{nm}=\sum_{k=0}^{d-1}e^{\frac{2i\pi nk}{d}}|k\oplus m\rangle\langle k|,
\end{eqnarray}
where $n,m = 0$ to $d-1$.
In the case for $d=2$, this set of unitary operators can be rewritten as
\begin{eqnarray}
 U_{00}=|0\rangle\langle 0|+|1\rangle\langle 1|,\nonumber\\
U_{01}=|0\rangle\langle 1|+|1\rangle\langle 0|,\nonumber\\
U_{10}=|0\rangle\langle 0|+e^{i\pi}|1\rangle\langle 1|,\nonumber\\
U_{11}=|0\rangle\langle 1|+e^{i\pi}|1\rangle\langle 0|.
\end{eqnarray}
Let us apply this set of linearly independent unitary operators $\{U_{00},U_{01},U_{10},U_{11}\}$ on the state $|\phi_{2}\rangle$ locally on one of its qubit. Then we have
\begin{eqnarray}
|\phi_{00}\rangle = (U_{00}\bigotimes I)|\phi_{2}\rangle=[C_0|0\rangle|0\rangle+C_1|1\rangle|1\rangle],\nonumber\\
|\phi_{01}\rangle = (U_{01}\bigotimes I)|\phi_{2}\rangle=[C_0|1\rangle|0\rangle+C_1|0\rangle|1\rangle],\nonumber\\
|\phi_{10}\rangle = (U_{10}\bigotimes I)|\phi_{2}\rangle=[C_0|0\rangle|0\rangle-C_1|1\rangle|1\rangle],\nonumber\\
|\phi_{11}\rangle = (U_{11}\bigotimes I)|\phi_{2}\rangle=[C_0|1\rangle|0\rangle-C_1|0\rangle|1\rangle].
\end{eqnarray}
In this set $\{|\phi_{00}\rangle,|\phi_{01}\rangle,|\phi_{10}\rangle,|\phi_{11}\rangle\}$ the states with different $m$ values are orthogonal to one another. But the states with same $m$ values need not be orthogonal. In any set of three states, at least two states will have the same $m$ value. So all of these three states are not going to be mutually orthogonal. So the pairs for $n=0$, $n=1$ are given by $\{|\phi_{00}\rangle,|\phi_{01}\rangle\}$, $\{|\phi_{10}\rangle,|\phi_{11}\rangle\}$.

Here we note that by taking a vector from each of the above classes one can also form orthogonal pairs like $\{|\phi_{00}\rangle,|\phi_{11}\rangle\}$, $\{|\phi_{10}\rangle,|\phi_{01}\rangle\}$. Even then one can not extend the dimension of the basis formed by these pairs. We see that one can have orthogonal basis $\{|\phi_{nm}\rangle,|\phi_{n^{'}m^{'}}\rangle\}$ for both $n=n^{'},n\neq n^{'}$. It is clearly evident that the orthogonality condition is independent of the value of $n$. The only requirement is to be a member of the pair must have different $m$ values $(m\neq m^{'})$.

Now, for a given pair,  there is no vector in the remaining set, which is orthogonal to the pair.
As an illustration, if we try to include $|\phi_{10}\rangle$ in the first pair   $\{|\phi_{00}\rangle,|\phi_{01}\rangle\}$,
it will be orthogonal to $|\phi_{01}\rangle$which has different $m$ value. However, it is not going to be 
orthogonal to $|\phi_{00}\rangle$ with the same $m$ value. This argument runs for every pair. This 
does not depend on the values of the index $n$. Thus, we see that the index $m$ actually discriminates the orthogonal vectors.
More formally, we show that indeed there exists no unitary transformation which will increase the cardinality of the set of the orthogonal vectors or in other words can extend the dimension of the basis. This shows that the basis is unextendible as far as the local unitaries are concerned.  To see this, let us consider the general unitary transformation
\begin{eqnarray}
 V=\sum_{p,q} f_{pq}U_{pq}.
\end{eqnarray}
Here $f_{pq}$ are complex coefficients  and $p,q$ vary from $0$ to $d-1$. In the above expression $U_{pq}$'s are unitary so is 
$V$. This in particular implies
 \begin{eqnarray}
\sum_{p,q} |f_{pq}|^2=1.  
 \end{eqnarray}
The action of the general unitary transformation on $|\phi_{2}\rangle$ is given by,
\begin{eqnarray}
 |\Phi\rangle=V|\phi_{2}\rangle= \sum_{p,q} f_{pq}U_{pq} |\phi_{2}\rangle\nonumber\\
=\sum_{p,q} f_{pq}\sum_kC_k e^{i\pi kp}|k\oplus q\rangle|k\rangle.
\end{eqnarray}
At this point, we would like to see whether the state $|\Phi\rangle$ is orthogonal to $\{|\phi_{00}\rangle,|\phi_{11}\rangle\}$ or not.  Without any loss of generality,  we have taken $n=n^{'}$ and in particular  $n=n^{'}=0$.  The argument also runs if $n=n^{'}=1$ and also for $n \neq n^{'}$.
If these states are orthogonal, then we will be able to find an additional state which is orthogonal to each of these two states. So in order to check that we consider the inner products
\begin{eqnarray}
 \langle\phi_{0m} |\Phi\rangle=\sum_p f_{pm}\sum_k |C_k|^2 e^{i\pi kp}.
\end{eqnarray}
For it to be zero for all $m$, we need  $\langle\phi_{00} |\Phi\rangle=0$ and $ \langle\phi_{01} |\Phi\rangle=0$. This implies that 
\begin{eqnarray}
f_{00}\sum_k |C_k|^2+f_{10}\sum_k |C_k|^2 e^{i\pi k}=0,\nonumber\\
f_{01}\sum_k |C_k|^2+f_{11}\sum_k |C_k|^2 e^{i\pi k}=0.
\end{eqnarray}
For this to be true we must have
\begin{eqnarray}
 f_{00}+f_{10}=0,f_{00}-f_{10}=0\Rightarrow f_{00}=f_{10}=0,\nonumber\\
f_{01}+f_{11}=0,f_{01}-f_{11}=0\Rightarrow f_{01}=f_{11}=0.
\end{eqnarray}
This implies  that all $f_{pq}= 0$ and hence $V=0$.
This clearly indicates  that there does not exist a  unitary transformation, that can increase the cardinality of the set of orthogonal
states $\{|\phi_{00}\rangle,|\phi_{01}\rangle\}$. Thus, we cannot extend the dimension of the basis formed by this set.

\section{Locally Unextendible Non-Maximally Entangled Basis in $H^3 \bigotimes H^3$ }

Let us take a non-maximally entangled state  $|\phi_{3}\rangle$ in $ H^3 \bigotimes H^3$ given by 
\begin{eqnarray}
 |\phi_{3}\rangle 
= C_0|0\rangle|0\rangle+C_1|1\rangle|1\rangle+C_2|2\rangle|2\rangle.
 \end{eqnarray}
The set of local unitary operators $\{U_{00},U_{01},U_{02},U_{10},U_{11},U_{12},U_{20},U_{21},U_{22}\}$ to be applied on  $|\phi_{3}\rangle$ are given in  equation (2).
Quite similar to the qubit case, on applying these local unitaries on $|\phi_{3}\rangle$, we will get the set of nine vectors $\{|\phi_{nm}\rangle,n,m=0,1,2\}$. Out of this set of nine vectors if we take any three of them independent of $n$,  having different $m$ values, we will find them to be mutually orthogonal. However, if we try to include more than three vectors in this subset, the mutual orthogonal property is no longer obeyed. Thus, none of the vectors in the complementary subspace spanned by these three vectors are orthogonal to all of these three vectors.
Mathematically, we can see this just like the qubit case.  We use the same general unitary transformation as defined in equation (5).
The action of this general unitary operator on $|\phi_{3}\rangle$ is given by
\begin{eqnarray}
 |\Psi\rangle=V|\phi_{3}\rangle= \sum_{p,q} f_{pq}U_{pq}|\phi_{3}\rangle\nonumber\\
=\sum_{p,q} f_{pq}\sum_kC_k e^{i\pi kp}|k\oplus q\rangle|k\rangle.
\end{eqnarray}
Next, we check whether the vector $ |\Psi\rangle$ is orthogonal to all the three vectors $|\phi_{0m}\rangle$ or not.  Here also without any loss of generality we have considered the case, when $n$ values of the members of the set are equal. In particular, here we have taken $n=0$. One can also take $n=1,2$. The argument holds true even when the $n$ values are no longer equal. 
On equating these inner products to zero, and on further simplification, the condition for having a vector orthogonal to all these three vectors boils down to the condition of having a non trivial solution to the following three simultaneous equations
\begin{eqnarray}
f_{00}+f_{10}+f_{20}=0,\nonumber\\
f_{00}-f_{10}+f_{20}=0,\nonumber\\
f_{00}+f_{10}-f_{20}=0.
\end{eqnarray}
Since the determinant value of the coefficient matrix of the above set of simultaneous equation is not equal to zero, so the only possible solution to these equations is the trivial solution. This implies that the most general unitary operator $V$ is a null operator. This proves that indeed in general there does not exist any unitary operator that can increase the dimension of the subspace spanned by the set of orthogonal states $\{|\phi_{00}\rangle,|\phi_{01}\rangle\,|\phi_{02}\rangle\}$. Thus, LUNMEB in $H^3 \bigotimes H^3$ is only of dimension three.

\section{Locally Unextendible Non-Maximally Entangled Basis in $H^d \bigotimes H^d$}

In this section, we construct locally unextendible non-maximally entangled basis (LUNMEB) for a $d\bigotimes d$ dimensional system.
In order to do that, we start with a non-maximally entangled state and then apply local unitaries to it. We consider two types of non-maximally entangled states. In the first subsection, we consider a genuinely non-maximally entangled state. In the second subsection, we consider non-maximally entangled state which is however maximally entangled in its subspace. In the first case we construct a LUNMEB of dimension $d$, while in the second case the basis is of dimension $(d-1)^2$.

\subsection{Non-Maximally Entangled states in $H^d\bigotimes H^d$}

Let us consider a non-maximally entangled set of vectors in $H^d \bigotimes H^d$.
The elements of this non-maximally entangled set of vectors  are given by,
\begin{equation}
 |\phi_{nm}\rangle=N_{nm}\sum_{j=0}^{d-1}c_j^{nm}|j\rangle|j\oplus m\rangle,
\end{equation}
where $N_{nm}$ is the normalization constant and is equal to $\frac{1}{\sqrt{\sum_{j=0}^{d-1}|c_j^{nm}|^2}}$. If the above set is part of a set of $d^2$ orthonormal
basis vectors, then the coefficients $c_j^{nm}$ should satisfy the
following condition,
\begin{equation}
N_{nm}N_{pm}\sum_{k=0}^{d-1}c_k^{*nm}c_k^{pm}=\delta_{np},
\end{equation}
where the indices $n$, $p$, $m$ and $k$ take integer values between $0$
and $d-1$.  For a system of two qudits, these vectors
$\{|\phi_{nm}\rangle\}$ naturally fall into $d$ classes. Each class is labeled
by $n$. Within each class, there are $d$ states, which are labeled
by $m$. We now prove the central result of this paper.

\textbf{Theorem:}  For a given non-maximally entangled state, we can construct at most $d$
mutually orthogonal states by applying local unitary transformations. This set of states is being
labeled as locally unextendible non-maximally entangled basis (LUNMEB).

\textbf{Proof:}
To prove this, we start with any one of the basis vector from the set  $\{|\phi_{nm}\rangle\}$ and apply a specific set of linearly
independent unitary operators to see that for a fixed value of $n$ there are only $d$ vectors which are orthogonal to each other. However, the question remains whether there exists any unitary transformation which will extend the dimension of this basis from $d$ to $d+1$. We find the answer to this question is no.
Let us consider a vector (qudit) from this set of vectors $\{|\phi_{nm}\rangle\}$, written explicitly in the Schmidt decomposition form as
\begin{equation}
 |\phi\rangle=\sum_{k=0}^{d-1}C_k|kk\rangle.
\end{equation}
Here we consider a specific set of linearly independent unitary operators $U_{nm}$ which itself form a
basis. This set of unitary operators are $U_{nm}=\sum_{k=0}^{d-1}e^{\frac{2i\pi nk}{d}}|k\oplus m\rangle\langle k|$.
The  action of these unitary operators on a specific state $|p\rangle$ is given by
\begin{equation}
 U_{nm}|p\rangle=\sum_{k=0}^{d-1}e^{\frac{2i\pi nk}{d}}|k\oplus m\rangle\langle k|p\rangle\nonumber\\
=e^{\frac{2i\pi np}{d}}|p\oplus m\rangle.
\end{equation}
Next we apply two different general unitary transformations $U_{nm}$, $U_{n^{'}m^{'}}$  on  $|\phi\rangle$  to see how many of these 
states could be orthogonal. After applying the unitary transformations the resultant states are given by
\begin{eqnarray}
 |\psi_{nm}\rangle=U_{nm}|\phi\rangle=\sum_{p=0}^{d-1}C_p e^{\frac{2i\pi np}{d}}|p\oplus m\rangle|p\rangle,\nonumber\\
|\psi_{n^{'}m^{'}}\rangle=U_{n^{'}m^{'}}|\phi\rangle=\sum_{q=0}^{d-1}C_q e^{\frac{2i\pi n^{'}q}{d}}|q\oplus m^{'}\rangle|q\rangle.
\end{eqnarray}
Here we would like to find out the range of $n,m,n^{'},m^{'}$, for which the states $|\psi_{nm}\rangle$ and $|\psi_{n^{'}m^{'}}\rangle$ 
are orthogonal to each other. Now, taking the inner product between these two vectors we obtain
\begin{eqnarray}
&&\langle \psi_{nm}|\psi_{n^{'}m^{'}}\rangle{}\nonumber\\&&=\sum_{p,q} C^{*}_pC_q e^{\frac{-2i\pi np}{d}} e^{\frac{2i\pi n^{'}q}{d}}\langle 
p\oplus m|q\oplus m^{'}\rangle\delta_{pq}
{}\nonumber\\&&=\sum_p |C_p|^2 e^{\frac{2i\pi (n^{'}-n)p}{d}}\langle p\oplus m|p\oplus m^{'}\rangle.
\end{eqnarray}
This implies that we have
\begin{eqnarray}
\delta_{nn^{'}}\delta_{mm^{'}}=\sum_p |C_p|^2 e^{\frac{2i\pi (n^{'}-n)p}{d}}\delta_{mm^{'}}\\
{\rm or}~~ \sum_p |C_p|^2 e^{\frac{2i\pi (n^{'}-n)p}{d}}=\delta_{nn^{'}}.
\end{eqnarray}
The condition obtained in equation (19) can be satisfied only when $m\neq m^{'}$. So the set of orthogonal states will have different $m$ values. The orthogonality condition is independent of $n$ and will hold for both $n= n^{'}$ and $n\neq n^{'}$. So, without any loss of generality, we built up the classes for fixing the values of $n$. Since $n$ runs from $0$ to $d-1$, we can have $d$ classes and for each class we will have $d$ orthogonal vectors as $m$  varies from $0$ to $d-1$. 
For a given $n$, this set is given by $\{|\psi_{n0}\rangle,|\psi_{n1}\rangle,.....,|\psi_{n(d-1)}\rangle\}$. So this set of basis vectors clearly satisfies the first three conditions of the definition of LUNMEB. 
Next we wish to show that if we apply most general unitary transformations on any of the basis state we cannot get an additional orthogonal state.
Let us consider the most general unitary transformation $V$ already defined in equation (5).
Without any loss of generality, let us choose $|\psi_{00}\rangle$ from the set $\{|\psi_{00}\rangle,.....,|\psi_{0(d-1)}\rangle\}$ 
and apply the most general unitary transformation. The resultant state after this transformation is given by 
\begin{eqnarray}
 |\Phi\rangle=V|\psi_{00}\rangle=\sum_{p,q} f_{pq}U_{pq}|\psi_{00}\rangle\nonumber\\
=\sum_{p,q} f_{pq}\sum_{k}C_{k} e^{\frac{2i\pi kp}{d}}|k\oplus q\rangle|k\rangle.
\end{eqnarray}
At this point we would like to see whether the state $|\Phi\rangle$ is orthogonal to $|\psi_{0m}\rangle$ or not. 
If it is, then we will be able to find an additional state which is orthogonal to each of these $d$ states. 
So in order to check that we consider the inner products
\begin{eqnarray}
 \langle \psi_{0m}|\Phi\rangle=\sum_{j} C_j^{*}\langle j\oplus m|\langle j|\Phi\rangle\nonumber\\
=\sum_{p,q} f_{pq}\sum_{k}|C_{k}|^2 e^{\frac{2i\pi kp}{d}}\langle k\oplus m|k\oplus q\rangle\nonumber\\
=\sum_{p} f_{pm}\sum_{k}|C_{k}|^2 e^{\frac{2i\pi kp}{d}}.
\end{eqnarray}
Next we would like to see whether the above expression can be zero for all $m$, i.e
$\sum_{p} f_{pm}\sum_{k}|C_{k}|^2 e^{\frac{2i\pi kp}{d}}=0, \forall m$. This implies that we have 
\begin{eqnarray}
\sum_{p} f_{pm}(|C_0|^2+e^{\frac{2i\pi p}{d}}|C_1|^2+....\nonumber\\+e^{\frac{2i\pi (d-1)p}{d}}|C_{d-1}|^2)=0. \forall m 
\end{eqnarray}
Since $|C_k|^2\neq 0$, this can only be zero, if we have the following set of equations
\begin{eqnarray}
 \sum_{p} f_{pm}=0,\nonumber\\\sum_{p} f_{pm} e^{\frac{2i\pi p}{d}}=0,\nonumber\\....,\nonumber\\\sum_{p} f_{pm} e^{\frac{2i\pi (d-1)p}{d}}=0.
\end{eqnarray}
One solution for this above set of equations is the trivial solution, i.e, 
$ f_{pm}=0,\forall m$. This implies that $ V=0$.
This shows that we cannot construct additional orthogonal vector.
The question still remains whether there exists a non trivial solution for this set of equations or not. 
We note that these conditions are independent of $m$. So if there is a non trivial solution for one $m$, then it will be true for all $m$. For a non trivial solution to exist we must have  $det(e^{\frac{2i\pi pk}{d}})=0$, where $p,k=0,1,...,(d-1)$. Since we know that $\frac{e^{\frac{2i\pi pk}{d}}}{\sqrt{d}}$  is an unitary matrix with a determinant value equal to 1, so this implies $det(e^{\frac{2i\pi pk}{d}})\neq 0$. Therefore, the only solution for the system of equations is the trivial solution, i.e $f_{pm}=0$. This clearly indicates that there exists no local unitary transformation which can extend the basis beyond this $d$ orthogonal vectors. We conclude that the set of of states $\{|\psi_{n0}\rangle,|\psi_{n1}\rangle,.....,|\psi_{n(d-1)}\rangle\}$ forms a LUNMEB in $H^d\bigotimes H^d$. 

\subsection{Maximally Entangled states in subspace of $H^d\bigotimes H^d$}

In this subsection, we consider the case when a non-maximally entangled set in $H^d\bigotimes H^d$ is maximally entangled in the subspace $H^{d-1}\bigotimes H^{d-1}$. Let us consider a maximally entangled state in this subspace which is given by
\begin{equation}
 |\phi^{'}\rangle=\frac{1}{\sqrt{d-1}}\sum_{k=0}^{d-2}|kk\rangle.
\end{equation}
Here we see that there exists a set of unitary operators which when acted on this particular state locally gives rise to a set of $(d-1)^2$ set of orthogonal basis vectors. We call this set of these basis vectors locally unextendible in the sense that there exists no vectors in the complementary subspace which will be orthogonal to all of these vectors.
The set of unitary operators which are going to serve our purpose is given by
\begin{eqnarray}
 U^{'}_{nm}=\sum_{k=0}^{d-2} e^{\frac{2i\pi nk}{d-1}}|k\oplus m\rangle\langle k|.
\end{eqnarray}
Quite similar to the previous subsection, here also we apply two general unitary transformations $U^{'}_{nm}$ and $U^{'}_{n^{'}m^{'}}$ on $|\phi^{'}\rangle$ to see how many of them are orthogonal to each other. Taking the inner products between the two resultant vectors  $|\psi^{'}_{nm}\rangle$ and $|\psi^{'}_{n^{'}m^{'}}\rangle$, we get the condition,
\begin{eqnarray}
\langle \psi^{'}_{nm}|\psi^{'}_{n^{'}m^{'}}\rangle=\frac{1}{d-1}\sum_{p=0}^{d-2} e^{\frac{2i\pi (n^{'}-n)p}{d-1}}\delta_{mm^{'}}.
\end{eqnarray}
This implies that we have 
\begin{eqnarray}
\delta_{nn^{'}}=\frac{1}{d-1}\sum_{p=0}^{d-2} e^{\frac{2i\pi (n^{'}-n)p}{d-1}}\nonumber\\\Rightarrow \delta_{nn^{'}}=\delta_{n-n^{'}0}.
\end{eqnarray}
This expression is true for all values of $n$ and $n^{'}$. Not only that it is also independent of $m$. So for each value of $n$ there are $m=0,1,....(d-2)$ orthogonal vectors and $n$ itself ranges from $0,1,...,(d-2)$. Therefore, together the basis contains $(d-1)^2$ vectors. Thus we see that  if we start with a non-maximally entangled state which is maximally entangled in its subspace and apply local unitaries we can generate a basis of dimension $(d-1)^2$. For example, in the case of $d=3,4$, the number of orthogonal basis vectors are $4,9$ respectively.
One can show this basis to be unextendible in the sense that if one applies the most general unitary transformation on $|\phi^{'}\rangle$ , the resultant state obtained is not orthogonal to all members of the basis.
Thus we see that in principle we are able to generate a LUNMEB in $H^d \bigotimes H^d$ of dimension $(d-1)^2$. This is only possible if the non-maximally entangled state to start with is maximally entangled in the largest subspace. 

\section{POVM for Unambiguous discrimination of non-maximally entangled state of vectors}

For general $d^2$ non-maximally entangled states of vectors we have seen that there are $d$ classes of vectors
each containing $d$ vectors which are mutually orthogonal to each other. So in principle we can distinguish these 
$d$ vectors. However, if we take one vector from each of these $d$ classes, they are not orthogonal to 
each other. It turns out that we can distinguish these states unambiguously, if not perfectly,  by constructing the appropriate POVM operators.

Let us consider the following $d$ non-orthogonal vectors each taken from $d$ different classes,
\begin{eqnarray}
|\psi_{l0}\rangle \equiv |\psi_l\rangle=\sum_{k=0}^{d-1}\sqrt{p_k}e^{\frac{2\pi  ilk}{d}}|kk\rangle,
\end{eqnarray}
where $p_k$ are the Schmidt coefficients and the index $l$ takes values from $0,1,...(d-1)$. 
Also $p_k = C_{k}^2$.
Our goal is not necessarily to construct optimal set of POVM operators, but just a set of operators
for unambiguous discrimination of these states. One technique to construct a set of such operators
is to find a orthogonal vector corresponding to each vector in such a way that it is not orthogonal to 
the remaining $d-1$ vectors. One such construction is
\begin{eqnarray}
|\bar{\psi_l}\rangle=N[-\frac{d-1}{\sqrt{p_0}}|00\rangle+\sum_{k=1}^{d-1}\frac{1}{\sqrt{p_k}} e^{-\frac{2\pi ilk}{d}}|kk\rangle], 
\end{eqnarray}
where $l=0,1,...(d-1)$ and $N$ is the normalization constant.

The POVM operators are then given by,
\begin{eqnarray}
&&P_0=A|\bar{\psi_0}\rangle\langle \bar{\psi_0}|,P_1=A|\bar{\psi_1}\rangle\langle \bar{\psi_1}|,...,{}\nonumber\\&& P_{d-1}=A|\bar{\psi_{d-1}}\rangle\langle \bar{\psi_{d-1}}|,
P_{E}=I-\sum_{i=0}^{d-1} P_i.
\end{eqnarray}
where $A$ is a constant and $P_{E}$ is the operator corresponding to the inconclusive outcome. We choose  the constant $A$ such that the operator $P_{E}$ is positive. Without any loss of generality, we assume the ordering of the
Schmidt's coefficients to be  $p_{d-1}>p_{d-2}>.....>p_{0}$.  Then we can choose $A$  as 
$\frac{p_0}{d(d-1)|N|^2}$ to make  $P_{E}$ positive.
Therefore the the probability of failure for not discriminating the states is given by, $P_{error}=\langle \psi_0|P_{E}
|\psi_0\rangle=p_1+p_2+...+p_{d-1}-\frac{p_0}{d-1}=1-p_0 \frac{d}{d-1}$.  The probability of the success 
$P_{success}=1-P_{error}=p_0\frac{d}{d-1}$.
Thus we show that in principle we can construct POVM operators which can distinguish these states with a certain probability of success.

   The probability of distinguishing $d$ non-maximally entangled states of a qudit has also been considered in Ref \cite{ppa}.
  However, there appropriate POVM operators were not constructed. Instead, the approach
  of Duan and Guo \cite{dg} was taken. Their method is based on finding an appropriate unitary operator 
  and post selection of measurement action.

\section{Applications in Superdense Coding}
Superdense coding is a technique used in quantum information theory to transmit classical information by sending
quantum systems \cite{ben3}. In the simplest case, Alice wants to send Bob a binary number $x \in \{00,01,10,11\}$. She picks up one of the unitary operators $\{I,X,Y,Z\}$ according to $x$ she has chosen and applies the transformation on her qubit (the first qubit of the Bell state shared by them). Alice sends her qubit to Bob after one of the local unitaries 
are applied. The state obtained by Bob will be one of the four basis vectors, so he performs the measurement in the Bell basis to obtain two bits of information. It is quite well known that if we have a maximally entangled state in $H^d\bigotimes H^d$ as our resource, then we can send $2\log d$ bits of classical information. In the asymptotic case, we know one can send $\log d+S(\rho)$ amount of bit  when one considers non-maximally entangled state as resource \cite{pati,hiro, bru,bru1, shad}. 
As an application of our result, we see that {\em any} entangled state is suitable for superdense coding. We have
seen that when Alice and Bob share a non-maximally entangled state, then  Alice can create $d^2$  vectors with the aid of local unitaries. Out of which these $d^2$ vectors we can create $d$ classes; each class containing d vectors which are  mutually orthogonal and thus forming an unextendible basis. In principle Bob will be able to distinguish $d$ orthogonal vectors. However,  Bob will not be able to distinguish perfectly the remaining vectors from these vectors as they are not mutually orthogonal. So Alice in principle can send $(1+p)\log d$ (where $p$ is the success probability of
distinguishing $d$ non-orthogonal states) bits of classical information. It has been seen in the previous section that by constructing the appropriate POVM operators we can have the success probability $p$ equals to $p_0\frac{d}{d-1}$, where $p_0 = C_{0}^2$ is the smallest
Schmidt coefficient. For this set of POVM operators, the total number of bits Alice can send to Bob is $(1+p_{0}\frac{d}{d-1})\log d$ which is
more than $\log d$ bits which can be sent without entanglement.
We also note that if we start with a non-maximally entangled state which is maximally entangled in a subspace of the original Hilbert space, then there is a set of local unitaries which will create $(d-1)^2$ orthogonal vectors. 
In this case, Alice can send at most $2\log_2 (d-1)$ bits.
 This is even true for the asymmetric cases like $H^{d_1}\bigotimes H^d$ and $H^d\bigotimes H^{d_1}$ (where $d_1<d$ and the maximal value $d_1$ can take is $d-1$)

  We can compare the maximum classical bits that can be sent using the state maximally entangled in the subspace
  with the state which is not. We see that $(1+p_{0}\frac{d}{d-1})\log d > 2\log_2 (d-1)$, if $p_0 > f_d$ where $f_d =({d-1 \over d \log d})\, \log({(d-1)^2 \over d})$. In such a situation, fully non-maximally entangled state would be more suitable for superdense coding. The function $f_d$ has been
  plotted in Fig 1. Since $p_0 < {1 \over d}$, we note that the fully non-maximally entangled state is more
  suitable for superdense coding when $ d \le 3$. For $d = 3$, the value of $p_0$ needs to be between 0.175 and 0.333. For $d=2$, any allowed
  value of $p_0$ will suffice. We note that our POVM operators may not be optimum, so it may be possible that a fully non-maximally state is  more suitable for superdense coding  even for $d > 3$.

\begin{figure}[t]
\begin{center}

\[
\begin{array}{cc}
\includegraphics[height=6.0cm,width=6cm]{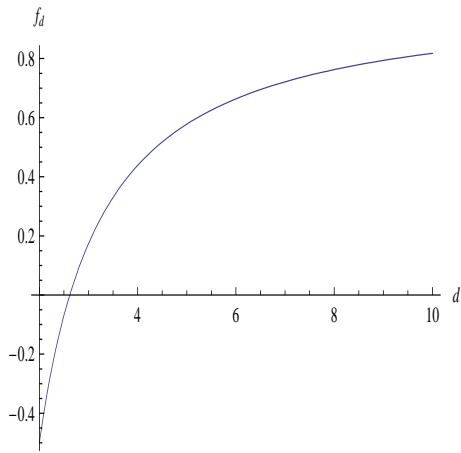}
\end{array}
\]
\caption{The function $f_d$ is plotted against the dimension $d$ of the basis}

\end{center}
\end{figure}
 
\section{Conclusion}

We have introduced the notion of locally unextendible non-maximally entangled basis (LUNEMB). They are unextendible in the sense that, there is no local unitary operator which will create a vector orthogonal to all members of the basis. We build up the work by constructing this set of basis vectors for $d=2$, $d=3$ and then generalizing it for arbitrary $d$. We find that such a basis in  $H^d\bigotimes H^d$ will have $d$ orthonormal vectors. We began with a genuinely non-maximally entangled state and applied a given set of unitary operators on one side of it. We find that out of the resultant vectors we can build up $d$ classes; each containing $d$ vectors. Each of these $d$ vectors are mutually orthogonal. We also showed that there
does not exist any unitary operator which can extend the dimension of the basis formed by this set of $d$ vectors. However,  if we consider a non- maximally entangled state which is maximally entangled in the sub space $H^{(d-1)}\bigotimes H^{(d-1)}$, then we can construct a basis which  has $(d-1)^2$ orthonormal vectors. This result has application
for superdense coding protocol. It shows that any entangled state can be used for superdense coding.
By explicitly constructing a set of POVM operators, we find  that Alice can send  at least $(1+p_0\frac{d}{d-1})\log d$
 (where $p_0$ is the smallest Schmidt's coefficient) bits of information to Bob. In the case of a maximally entangled state in the subspace, Alice can send at most $2 \log (d-1)$ bits of information. We also find that for $d = 3$ we can send
more classical bits with a fully non-maximally entangled state than with the state maximally entangled in the subspace.
This happens when the smallest Schmidt coefficient $p_0$ lies between  0.175 and 0.333.

\textit{Acknowledgment}: I.C acknowledges Dr. Satyabrata Adhikari for providing useful comments in the up gradation of manuscript.

\end{document}